\documentclass[aps,twocolumn,preprintnumbers,amsmath,amssymb,nofootinbib,superscriptaddress,notitlepage,10pt]{revtex4-1}

\usepackage{color}
\usepackage{epsfig}
\usepackage{txfonts}

\newcommand{\mpv}[1]{{#1}} 

\newcommand{\parag}[1]{\vspace{0.15cm} \paragraph*{\bf {#1}:}}

\begin{document}

\title{Heavy-hadron molecular spectrum from light-meson exchange saturation}

\author{Fang-Zheng Peng}
\affiliation{School of Physics,  Beihang University, Beijing 100191, China}

\author{Mario S\'anchez S\'anchez}
\affiliation{Centre d'\'Etudes Nucl\'eaires, CNRS/IN2P3, Universit\'e de Bordeaux, 33175 Gradignan, France}

\author{Mao-Jun Yan}
\affiliation{School of Physics,  Beihang University, Beijing 100191, China}

\author{Manuel Pavon Valderrama}\email{mpavon@buaa.edu.cn}
\affiliation{School of Physics,  Beihang University, Beijing 100191, China}

\date{\today}


\begin{abstract} 
  \rule{0ex}{3ex}
  If known, the spectrum of heavy-hadron molecules will be a key tool
  to disentangle the nature of the exotic states that are being
  discovered in experiments.
  Here we argue that the general features of the molecular spectrum
  can be deduced from the idea that the short-range interaction
  between the heavy hadrons is effectively described by scalar
  and vector meson exchange, i.e. the $\sigma$, $\omega$
  and $\rho$ mesons.
  By means of a contact-range theory where the couplings are saturated
  by the aforementioned light mesons we are indeed able to postdict
  {the $X(3872)$ (as a $D^* \bar{D}$ molecule) from
    the three $P_c(4312)$, $P_c(4440)$ and $P_c(4457)$ pentaquarks
    (as $\bar{D}\Sigma_c$ and $\bar{D}^* \Sigma_c$ molecules).}
  We predict a \mpv{$J^{PC} = 1^{--}$ $D \bar{D}_1$ molecule} at
  $4240-4260\,{\rm MeV}$ which might support the hypothesis
  that the $Y(4260)$ is at least partly molecular.
  The extension of these ideas to the light baryons requires minor
  modifications, after which we recover approximate SU(4)-Wigner symmetry
  in the two-nucleon system and approximately reproduce
  the masses of the deuteron and the virtual state.
\end{abstract}

\maketitle

\parag{1. Introduction}

Theoretical predictions of the hadronic spectrum are fundamental
for testing our understanding of strong interactions
against experiments.
SU(3)-flavor symmetry~\cite{Ne'eman:1961cd,GellMann:1962xb},
the quark model~\cite{Isgur:1978xj,Isgur:1978wd,Isgur:1979be,Godfrey:1985xj} and the theory behind quarkonium~\cite{Eichten:1978tg,Eichten:1979ms,Brambilla:1999xf,Brambilla:2004jw}
have provided valuable insights and clear predictions about
which hadrons to expect and their approximate masses.
With the experimental observation of exotic hadrons --- hadrons that are neither
three-quark states, a quark-antiquark pair or that do not fit into
preexisting quark-model predictions~\cite{Chen:2016qju,Hosaka:2016pey,Lebed:2016hpi,Guo:2017jvc} ---
new theoretical explanations have appeared,
among which molecular hadrons~\cite{Voloshin:1976ap,DeRujula:1976qd}
are popular.
However the spectrum of molecular hadrons is not properly understood,
with most theoretical applications of this idea being explicitly
customized to explain a particular hadron or a few at most.

The present manuscript attempts to overcome this limitation
by proposing a more general explanation of the spectrum of
hadronic molecules (in the spirit of
Refs.~\cite{Karliner:2015ina,Dong:2021juy}).
The idea is as follows: first, we will describe the interaction
between two heavy-hadrons in terms of a contact-range potential.
If the range of the binding mechanism between two heavy-hadrons is indeed
shorter than the size of the molecular state formed
by the aforementioned hadrons, a contact-range theory
will represent a good description.
Second, we will assume that the couplings in the contact-range potential
are saturated by light-meson exchange (i.e. $\sigma$, $\rho$ and $\omega$)
within the saturation procedure of Ref.~\cite{Peng:2020xrf}.
Third, for effectively combining the contribution from the saturation of
scalar and vector meson exchange, which happen at a different
renormalization scale as the masses of these light-mesons
are different, we will follow a renormalization group
equation (RGE).
This RGE will tell us what is the importance of scalar and vector
meson contributions to saturation relative to each other.
Fourth, the couplings derived from saturation are expected to be valid modulo a
proportionality constant, which we fix by solving the bound state equation
for a molecular candidate, for instance the $P_c(4312)$.
Finally, we can derive the predictions of this procedure
for other molecular states.

\parag{2. Saturation}

We will consider a generic two heavy-hadron system $H_1 H_2$, 
with $H_i = D^{(*)}$, $\Sigma_c^{(*)}$, etc. for $i=1,2$.
Owing to heavy-quark spin symmetry (HQSS) their interaction only depends
on the light-quark spins of heavy-hadrons $H_1$ and $H_2$,
which we denote $\vec{S}_{L1}$ and $\vec{S}_{L2}$.
If parity and light-spin are conserved in each vertex
(i.e. the $H_1 H_2 \to H_2 H_1$ transition does not happen),
we can describe the $H_1 H_2$ system with a contact-range interaction
that admits the multipolar expansion:
\begin{eqnarray}
  V_C = C_0 + C_1\,\hat{\vec{S}}_{L1} \cdot \hat{\vec{S}}_{L2} +
  \dots \, , \label{eq:V-cont}
\end{eqnarray}
plus higher order terms, if present, with no momentum/energy dependence
and where $\hat{\vec{S}}_{Li} = \vec{S}_{Li} / S_{Li}$ is a reduced spin
operator for hadron $i=1,2$.
Higher multipoles can be built analogously but in practice the monopolar and
dipolar terms are more than enough to effectively describe the short-range
interaction in most molecules,
as higher multipoles will be suppressed~\cite{Peng:2020xrf}.
By using a contact theory we are assuming that pion dynamics
and coupled channel effects are perturbative
corrections~\cite{Valderrama:2012jv,Lu:2017dvm}
\mpv{and that the resulting two-body bound state is not compact enough
  as to resolve the short-distance details of
  the light-meson exchanges binding it.}
Finally, this contact-range potential has to be regularized,
for which a regulator function depending on a cutoff $\Lambda$ is introduced
and where the couplings become functions of this cutoff,
i.e. $C_J = C_J(\Lambda)$ for $J=0,1$.
Concrete regularization details will be discussed later.

To determine the $C_0$ and $C_1$ couplings we assume that they are
saturated with scalar- and vector-meson exchange.
We begin by writing the Lagrangians for light-meson exchange
in a suitable notation in which instead of the full heavy-hadron fields
we use effective non-relativistic fields with the quantum numbers of
the light-quarks within the hadrons~\cite{Valderrama:2019sid,Peng:2020xrf}.
This is motivated by the observation that in heavy-hadron interactions
the heavy-quarks are spectators.
For the interaction of a scalar meson with the light-quark degrees of freedom
inside a heavy hadron, the Lagrangian reads
\begin{eqnarray}
  \mathcal{L} = g_{S} \, {q}_L^{\dagger} \sigma q_L \, ,
\end{eqnarray}
where $g_{S}$ is a coupling constant, $\sigma$ the scalar meson field
and $q_L$ the aforementioned effective non-relativistic
{\it light-quark subfield}.
For the vector mesons the Lagrangian can be written as a multipole expansion
\begin{eqnarray}
  \mathcal{L} &=& \mathcal{L}_{E0} + \mathcal{L}_{M1} + \dots
  \nonumber \\
  &=& g_{V} \, {q}_L^{\dagger} V_0 q_L  
  + \frac{f_{V}}{2 M} \, {q_L}^{\dagger} \,
  \epsilon_{ijk} \hat{\vec{S}}_{L i} \partial_j V_k \, q_L  + \dots \, ,
  \label{eq:lagrangian-vector}
\end{eqnarray}
where the dots indicate higher order multipoles.
In this Lagrangian, $g_V$ and $f_V$ are coupling constants,
$\epsilon_{ijk}$ the Levi-Civita symbol,
$V_{\mu} = (V_0, \vec{V})$ is the vector-meson field and $M$ is
the characteristic mass scale for this multipolar expansion
which for convenience we will set to be the nucleon mass,
$M = m_N \simeq 938.9\,{\rm MeV}$.
For simplicity we are not writing down explicitly
the isospin or flavor indices.
The number of terms depends on the spin of the light quark degrees of
freedom, where for $S_L = 0$ ($\Lambda_c$) there is only the electric term,
for $S_L = \frac{1}{2}$  ($D^{(*)}$)
there is also the magnetic dipole term,
for $S_L = 1$ ($\Sigma_c^{(*)}$)
an electric quadrupole term (which we will ignore), and so on.

In the $H_1 H_2$ system the potential for the scalar meson is
\begin{eqnarray}
    V_{S}(\vec{q}) &=& -\frac{g_{S1} g_{S2}}{{\vec{q}\,}^2 + m_{S}^2} \, ,
\end{eqnarray}
with $\vec{q}$ the exchanged momentum and where $g_{Si}$ refers to
the scalar coupling for hadron $i=1,2$,
while for the vector mesons we have
\begin{eqnarray}
  V_{E0}(\vec{q}) &=&
  +\frac{g_{V1} g_{V2}}{{\vec{q}\,}^2 + m_V^2} \, , \label{eq:E0} \\
  V_{M1}(\vec{q})
  &=& +\frac{2}{3}\,\frac{f_{V1} f_{V2}}{4 M^2}\,
  \hat{\vec{S}}_{L1} \cdot \hat{\vec{S}}_{L2}\,
  \frac{m_V^2}{{\vec{q}\,}^2 + m_V^2} + \dots \, , \label{eq:M1}
\end{eqnarray}
with $g_{Vi}$, $f_{Vi}$ the couplings for hadron $i=1,2$ and where
the dots indicate terms that vanish for S-wave or contact-range terms
for which the range is shorter than vector-meson exchange.
Following Ref.~\cite{Peng:2020xrf}, the saturation condition  
for scalar-meson exchange reads
\begin{eqnarray}
  C_0^S(\Lambda \sim m_{S}) &\propto& - \frac{g_{S1} g_{S2}}{m_{S}^2} \, ,
\end{eqnarray}
where we stress that saturation is expected to work for $\Lambda$
of the same order of magnitude as the mass of the exchanged
light-meson (thus $\Lambda \sim m_S$).
For vector-meson exchange we have
\begin{eqnarray}
  C_0^V(\Lambda \sim m_V) &\propto& \frac{g_{V1} g_{V2}}{m_V^2}\,\left[
    \zeta + \hat{\vec{I}}_1 \cdot \hat{\vec{I}}_2 \right] \, , \\
  C_1^V(\Lambda \sim m_V) &\propto& \frac{f_{V1} f_{V2}}{6 M^2}\,\left[
    \zeta + \hat{\vec{I}}_1 \cdot \hat{\vec{I}}_2 \right] \, ,
  \label{eq:C1-sat}
\end{eqnarray}
where we have now included isospin explicitly,
$\zeta = \pm 1$ is a sign to indicate the contribution
from the omega ($+1$ for $DD$, $\Sigma_c \bar{D}$, $\Sigma_c \Sigma_c$
and $-1$ for $D \bar{D}$, $\Sigma_c {D}$, $\Sigma_c \bar{\Sigma}_c$)
and $\hat{\vec{I}}_i = \vec{I}_i/I_{i}$ are normalized isospin
operators for the rho contribution (with $\vec{I}_i$ the standard
isospin operator and $I_{i}$ the isospin of hadron $i=1,2$).

\parag{3. Renormalization Group Evolution}

The saturation of the $C_0$ coupling receives contributions from two types
of light-mesons with different masses.
To combine the saturation from scalar- and vector-meson
exchange into a single coupling, we have to know first
the RGE of the couplings,
which for non-relativistic contact-range theories is well-understood and
follows the equation~\cite{Valderrama:2014vra,Valderrama:2016koj}
\begin{eqnarray}
  \frac{d}{d \Lambda} \left[ \frac{C(\Lambda)}{\Lambda^\alpha} \right] = 0 \quad
  \mbox{or, equivalently: } \,
  \frac{C(\Lambda_1)}{\Lambda_1^\alpha} =
  \frac{C(\Lambda_2)}{\Lambda_2^\alpha} \, , \quad
  \label{eq:RGE}
\end{eqnarray}
with $\alpha$ the anomalous dimension of the coupling.
From this we can combine the scalar- and vector-meson contributions as
\begin{eqnarray}
  C^{\rm sat}(m_V) =
  C^V(m_V) + {\left(\frac{m_V}{m_S}\right)}^{\alpha} C^S(m_S) \, .
  \label{eq:coupling-combo}
\end{eqnarray}
The anomalous dimension is linked with the behavior of the two-body
wavefunction $\Psi(R)$ at distances $R^{-1} \sim \Lambda$ by
${| \Psi(R \sim 1/\Lambda) |}^2 \sim \Lambda^{-\alpha}$~\cite{Valderrama:2014vra,Valderrama:2016koj} (i.e. $\alpha$ encodes the short-range
suppression of the wavefunction).
We do not know the exact form of the short-distance wavefunction,
but owing to the large mass of the heavy-hadrons it is sensible to assume
that the semiclassical approximation applies.
From including the Langer correction~\cite{Langer:1937qr}
we estimate $\Psi (R) \sim R^{1/2}$
which implies $\alpha = +1$. 
We end up with:
\begin{eqnarray}
  && C^{\rm sat}(\Lambda = m_V) =
  C_0^{\rm sat} + C_1^{\rm sat}\,\hat{\vec{S}}_{L1} \cdot \hat{\vec{S}}_{L2}
  \nonumber \\
  && \quad \propto 
  g_{V1} g_{V2}
  \left[ \zeta + \hat{T}_{12} \right]
  \left( \frac{1}{m_V^2} + \frac{\kappa_{V1} \kappa_{V2}}{6 M^2}\,\hat{C}_{L12} \right)
    - {(\frac{m_V}{m_S})}^{\alpha}\frac{g_{S1} g_{S2}}{m_S^2} \, , \nonumber \\
    \label{eq:coupling-sat}
\end{eqnarray}
where $\hat{T}_{12} = \hat{\vec{I}}_1 \cdot \hat{\vec{I}}_2$,
$\hat{C}_{L12} = \hat{\vec{S}}_{L1} \cdot \hat{\vec{S}}_{L2}$
and $\kappa_{Vi} = f_{Vi} / g_{Vi}$.
However we do not know (yet) the proportionality constant.
\mpv{
  It is worth stressing that
  the previous relations implicitly assume that the contact-range
  interaction has indeed been regularized and that the regularization scale
  (i.e. the cutoff $\Lambda$) is of the right order of magnitude
  ($\Lambda \sim m_V$). In the next lines we will explain
  in more detail our regularization prescription.
}

\parag{4. Predictions}

We now explain how to make predictions with the saturated coupling
and overcome the ambiguities in its exact definition.
First, we regularize the potential
\begin{eqnarray}
  \langle p' | V_C | p \rangle = C^{\rm sat}_{\rm mol}(\Lambda_H)\,
  f(\frac{p'}{\Lambda_H})\,f(\frac{p}{\Lambda_{H}}) \, ,
\end{eqnarray}
where $C^{\rm sat}_{\rm mol}$ is the saturated coupling
of Eq.~(\ref{eq:coupling-sat}) particularized for
a given molecule, $f(x)$ is a regulator function \mpv{(for which we choose
  a Gaussian regulator, $f(x) = e^{-x^2}$)} and
$\Lambda_{H}$ a {\it physical} cutoff, i.e. a cutoff which corresponds
with the natural hadronic momentum scale.
The cutoff $\Lambda_H$ can be either the scale at which saturation
is expected to work (from $m_S$ to $m_V$) or the momentum scale
at which we begin to see the internal structure of
the heavy-hadrons, \mpv{as these two scales are of
  the same order of magnitude}.
We opt for the later: $\Lambda_H = 1\,{\rm GeV}$.
This contact-range potential can be introduced within a bound state
equation to make predictions:
\begin{eqnarray}
  1 + 2\mu_{\rm mol}\,C^{\rm sat}_{\rm mol}(\Lambda_H)\,
  \int_0^{\infty} \frac{p^2\,d p}{2 \pi^2} \frac{f(p/\Lambda_H)^2}
  {p^2 + \gamma_{\rm mol}^2} = 0 \, , \label{eq:bound}
\end{eqnarray}
where $\mu_{\rm mol}$ is the reduced mass of the two-hadron system under
consideration and $\gamma_{\rm mol}$ the wave number of the bound state,
related to the binding energy by
$B_{\rm mol} = - \gamma_{\rm mol}^2 / 2 \mu_{\rm mol}$.
The mass of the predicted molecular state will be
$M_{\rm mol} = M_{\rm thres} - B_{\rm mol}$, with
$M_{\rm thres} = M_1 + M_2$ the threshold of the two-hadron system and
$M_i$ the mass of hadron $i=1,2$.

The proportionality constant between the contact-range coupling and
the saturation ansatz can be determined from a known molecular candidate.
Actually, the strength of the interaction is dependent
on the reduced mass times the coupling, $\mu_{\rm mol} C_{\rm mol}^{\rm sat}$.
If we take the $P_c(4312)$ as the reference molecule~\footnote{We notice
  that there is no ideal choice of a reference molecule, as for
  no exotic state there exists a clear consensus of
  its molecular nature.}
(${\rm mol} = P_c$),
we can first determine its coupling from solving Eq.~(\ref{eq:bound})
for this system and then define the ratio
\begin{eqnarray}
  R_{\rm mol} = \frac{\mu_{\rm mol} \, C_{\rm mol}^{\rm sat}}
  {\mu_{P_c} \, C_{P_c}^{\rm sat}} \, ,
  \label{eq:R-mol}
\end{eqnarray}
from which we can determine the interaction strength of a particular
molecule relative to the $P_c(4312)$.
Finally we plug $R_{\rm mol}$ and $C^{\rm sat}_{P_c}$ into Eq.~(\ref{eq:bound}):
\begin{eqnarray}
  1 + \left( 2\mu_{P_c}\,C^{\rm sat}_{P_c} \right)\,R_{\rm mol}\,
  \int_0^{\infty} \frac{p^2\,d p}{2 \pi^2} \frac{f(p/\Lambda_H)^2}
  {p^2 + \gamma_{\rm mol}^2} = 0 \, , \label{eq:bound-R}
\end{eqnarray}
and compute $\gamma_{\rm mol}$ and $B_{\rm mol}$ for a particular molecule.

For this we need $g_V$, $\kappa_V$ ($=f_V/g_V$) and $g_S$
for the heavy hadrons;
$g_V$ and $\kappa_V$ are determined from the mixing of the neutral vector mesons
with the electromagnetic current (i.e. Sakurai's universality and vector meson
dominance~\cite{Sakurai:1960ju,Kawarabayashi:1966kd,Riazuddin:1966sw}),
for which we apply the substitution rules
\begin{eqnarray}
\rho^0_{\mu} \to \frac{e}{2 g}\,A_{\mu} \quad \mbox{and} \quad
\omega_{\mu} \to \frac{e}{6 g}\,A_{\mu} \, ,
\label{eq:vector-em-mixing}
\end{eqnarray}
to Eq.~(\ref{eq:lagrangian-vector}) and match to
the light-quark contribution to the electromagnetic Lagrangian
($\rho^0_{\mu}$ and $\omega_{\mu}$ are the neutral rho and omega fields,
$A_{\mu}$ the photon field, $\mu$ a Lorentz index, $e$ the proton charge
and $g = m_V / 2 f_{\pi} \simeq 2.9$ the universal vector meson
coupling constant, with $m_V$ the vector meson mass and
$f_{\pi} \simeq 132\,{\rm MeV}$ the pion weak decay constant).
For $g_V$ we obtain $g_V = g$ ($2 g$) for $D^{(*)}$/$D_{0(1)}^{(*)}$/$D_{1(2)}^{(*)}$
($\Sigma_c^{(*)}$); 
$\kappa_V$ is proportional to the (light-quark) magnetic moment of
the heavy hadrons, which we calculate in the non-relativistic
quark-model~\cite{Riska:2000gd}, yielding
$\kappa_V = \tfrac{3}{2}\,({\mu_u}/{\mu_N})$ for $D^{(*)}$ and $\Sigma_c^{(*)}$,
$\kappa_V = \tfrac{3}{2}\,({\mu_u}/3 {\mu_N})$
($\tfrac{3}{2}\,(2 {\mu_u}/{\mu_N})$) for the $S^P_L = \tfrac{1}{2}^-$
($\tfrac{3}{2}^{-}$) P-wave charmed mesons $D_{0(1)}^{(*)}$ ($D_{1(2)}^{(*)}$)
where $\mu_N$ is the nuclear magneton and
$\mu_u \simeq 1.9\,\mu_N$ the magnetic moment of a constituent u-quark
\mpv{(for a more detailed account on the choice of the magnetic-like couplings,
we refer to Appendix \ref{app:couplings}).}
For $g_S$ we invoke the linear-$\sigma$ model~\cite{GellMann:1960np}
and the quark model~\cite{Riska:2000gd}, yielding
$g_S \simeq 3.4$ ($6.8$) for $D^{(*)}$ ($\Sigma_c^{(*)}$).
For the light-meson masses, we take $m_S = 475\,{\rm MeV}$ (the value in the
middle of the Review of Particle Physics (RPP) range of
$400-550\,{\rm MeV}$~\cite{Zyla:2020zbs}) and
$m_V = (m_{\rho} + m_{\omega})/2$ (with
$m_{\rho} = 770\,{\rm MeV}$ and
$m_{\omega} = 780\,{\rm MeV}$).

{
  At this point we find it worth mentioning
  that the light-meson exchange picture
  is not free of theoretical difficulties, the most important of which is
  probably the nature and width of the $\sigma$.
  This happens to be a well-known issue for which we briefly review a few of
  the available solutions in Appendix \ref{app:OBE}.
  Here it is enough to comment that when a broad meson is exchanged, it
  can be effectively approximated by a narrow one by a suitable
  redefinition of its parameters~\cite{Machleidt:1987hj,Machleidt:1989tm}.
  As we are determining the couplings from phenomenological relations
  which do not take into account the width of the scalar meson and
  provide a good description of a few molecular candidates,
  we consider this redefinition to have already taken place
  (we refer to Appendix \ref{app:OBE} for further discussion and details).
}

{
  Regarding uncertainties, the RGE indicates that the value of the saturated
  couplings are dominated by the scalar meson
  (see Eq.~(\ref{eq:coupling-combo})),
  which also happens to be the meson
  for which theoretical uncertainties are larger.
  For this reason we will generate error bands from the uncertainty
  in the scalar meson mass, $m_S = 475 \pm 75 \,{\rm MeV}$.
  Besides, the contact-range theory approximation also entails uncertainties:
  if the molecule is compact enough, the heavy hadrons will be able
  to resolve the details of the interaction binding them and
  the contact-range approximation will cease to be valid.
  We include a relative uncertainty of $\gamma_{\rm mol } / m_V$ (i.e. the ratio
  of the characteristic molecular momentum scale and the mass of
  the vector meson) to take into account this effect,
  which we then sum in quadrature with the previous error coming from $m_S$.
}

With these couplings we now reproduce
the $P_c(4312)$ with Eq.~(\ref{eq:bound}) yielding
$C_{P_c}^{\rm sat} = -0.80 \,{\rm fm}^2$ for $\Lambda_H = 1\,{\rm GeV}$.
After this we calculate $R_{\rm mol}$ and solve Eq.~(\ref{eq:bound-R})
to predict the molecules we show in Table~\ref{tab:predictions}:

\begin{enumerate}
\item[(i)] For molecules in the lowest isospin state the general pattern
  is that mass decreases with light-spin~\cite{Peng:2020xrf}
  (i.e. opposite to compact hadrons).

\item[(ii)] The origin of the light-spin dependence is the magnetic-like
  vector-meson exchange term in Eqs.~(\ref{eq:M1}) and (\ref{eq:C1-sat}).

\item[(iii)]
  For $\bar{D}^{(*)} \Sigma_c^{(*)}$, we reproduce
  the $P_c(4440$/$4457)$~\cite{Aaij:2019vzc} and
  find the spectrum predicted in Refs.~\cite{Liu:2019tjn,Valderrama:2019chc,Xiao:2019aya,Du:2019pij,Liu:2019zvb}.

\item[(iv)] For $D^{(*)} \bar{D}^{(*)}$,
  besides the $X$ and its $J^{PC} = 2^{++}$
  partner~\cite{Liu:2019stu,Tornqvist:1993ng,Valderrama:2012jv,Nieves:2012tt},
  the only other configuration close to binding is
  the $J^{PC} = 0^{++}$ $D \bar{D}$ system~\cite{Gamermann:2006nm,Xiao:2012iq}.
  It should be noticed that:
  \begin{itemize}
  \item[(iv.a)] The isoscalar $D^{(*)} \bar{D}^{(*)}$ systems can mix with
    nearby charmonia. This is a factor that we have not included
    in our model, yet discrepancies between our predictions and
    candidate states could point towards the existence of
    such mixing.
  \item[(iv.b)] In particular, the $X(3872)$ is predicted around
    its experimental mass, though with moderate uncertainties
    which do not exclude (but do not require either)
    a charmonium component.
    Though its nature is still
    debated~\cite{Swanson:2004pp,Dong:2009uf,Guo:2014taa,Esposito:2020ywk,Braaten:2020iqw},
    theoretical works point out to the existence of a non-trivial
    non-molecular component (e.g. Ref.~\cite{Esposito:2021vhu} finds
    a negative effective range in the $J^{PC} = 1^{++}$ $D^*\bar{D}$ system,
    which is difficult to explain in a purely molecular picture).
    The previous could be further confirmed if the spectroscopic
    uncertainties of the present model could be reduced
    to the point of determining if underbinding or
    overbinding exist.
  \item[(iv.c)] Had we used the $X(3872)$ as the reference molecule,
    the results of Table~\ref{tab:predictions} would have been almost
    identical to the ones we obtain from the $P_c(4312)$.
  \end{itemize}

\item[(v)]
  Molecules involving P-wave $D_1$/$D_2^*$ mesons have larger hyperfine
  splittings than their $D$/$D^*$ counterparts:
\begin{itemize}
\item[(v.a)] For instance, the $D_{1(2)}^{(*)} \bar{D}_{1(2)}^{(*)}$ molecules
  only bind for configurations with high light-spin, while the configurations
  with low light-spin content can even become repulsive.
\item[(v.b)] The hyperfine splitting of the $\Sigma_c \bar{D}_1$ pentaquarks
  is predicted to be approximately twice as large
  as for the $\Sigma_c \bar{D}^*$ one,
  about $30\,{\rm MeV}$ instead of $15\,{\rm MeV}$.
\end{itemize}

\item[(vi)]
  Molecules for which rho- and omega-exchange cancel out
  ($Z_c(3900)$~\cite{Ablikim:2020hsk}) or involving strangeness
  ($P_{cs}(4459)$~\cite{Aaij:2020gdg}) require
  additional discussion and are not listed.
  We advance that:
  \begin{itemize}
  \item[(vi.a)] 
    For the $Z_c$~\cite{Yang:2020nrt,Meng:2020ihj,Sun:2020hjw,Ikeno:2020csu}
    $I^G(J^{PC}) = 1^{+}(1^{+-})$ $D^*\bar{D}$ configuration, we predict
    a virtual state at $M = 3849.6\,{\rm MeV}$.
    This agrees with the EFT analysis of Ref.~\cite{Albaladejo:2015lob},
    which suggests that if the $Z_c$ is a virtual state its mass
    would be in the $M = (3831-3844)\,{\rm MeV}$ range,
    though with large uncertainties.
    Meanwhile in the EFT approach of Ref.~\cite{Yang:2020nrt}
    the $Z_c$ is located at $(3867-3871)\,{\rm MeV}$ when a virtual state
    solution is assumed ({\it fit 1} in Ref.~\cite{Yang:2020nrt}). 
    Recently Ref.~\cite{Yan:2021tcp} has proposed the inclusion of axial meson
    ($a_1(1260)$) exchange to explain the $Z_c$, in which case we would end up
    with a mass in the $(3856-3867)\,{\rm MeV}$ range.
    However this depends on the assumption that Eq.~(\ref{eq:coupling-combo})
    holds far away from $\Lambda = m_V$, which might very well not be correct
    without suitable modifications (check the discussion
    in Appendix \ref{app:OBE}).
    For comparison, Ref.~\cite{Dong:2021juy} explains the $Z_c$ states
    in terms of vector charmonia exchanges.
  \item[(vi.b)] If the scalar meson couples to $q=u,d,s$ with similar
  strength~\cite{Yan:2021tcp}, this will generate
  $P_{cs}$-like~\cite{Chen:2020uif,Peng:2020hql,Chen:2020kco,Liu:2020hcv}
  $I=0$, $J=\tfrac{1}{2}$, $\tfrac{3}{2}$ $\Xi_c \bar{D}^*$
  bound states at $M = 4466.9\,{\rm MeV}$.
  \end{itemize}

\item[(vii)] We remind that the results of Table~\ref{tab:predictions}
  ignore pion exchanges and coupled channel effects, which are
  considered to be perturbative corrections.
  Appendices \ref{app:OPE} and \ref{app:CC} explicitly check
  these two assumptions in a few concrete cases, leading in general
  to corrections that are indeed smaller than the uncertainties
  shown in Table~\ref{tab:predictions}.
  We notice that there might be specific molecules
  for which these assumptions do not hold.
\end{enumerate}

Finally we warn that though the formalism is identical to the one used
in a typical contact-range effective field theory (EFT),
this is not EFT: here the cutoff is not {\it auxiliary},
but physical.
It is not expected to run freely but a parameter chosen to reproduce
the known spectrum.

\begin{table}[!ttt]
\begin{tabular}{|ccccccc|}
\hline\hline
System  & $I$($J^{P(C)}$) & $R_{\rm mol}$ & $B_{\rm mol}$ & $M_{\rm mol}$ &
Candidate & $M_{\rm candidate}$\\
\hline
$np$ & $0$ $(1^+)$ & $1.66^{+0.12}_{-0.14}$ & $0.9^{+1.2}_{-0.8}$ & $1876.9^{+0.8}_{-1.2}$ & deuteron & $1875.6$ \\
$np$ & $1$ $(0^+)$ & $1.54^{+0.16}_{-0.20}$ & $0.2^{+1.1}_{-0.4}$ & $1877.7^{+0.2}_{-1.1}$ & $^1S_0$ pole  & $1877.8$ \\
\hline \hline
$N D$ & $0$ ($\tfrac{1}{2}^-$) & $1.01^{+0.06}_{-0.05}$ & $(0.7^{+0.6}_{-0.5})^V$ & $2805.5^{+0.5}_{-0.6}$ & $\Lambda_c(2765)$ & $2766.6$ \\
$N D^*$ & $0$ ($\tfrac{3}{2}^-$) & $1.19^{+0.12}_{-0.11}$ & $0.1^{+0.9}_{-0.2}$ & $2947.4^{+0.1}_{-0.9}$ & $\Lambda_c(2940)$ & $2939.6$ \\ 
\hline
$N D$ & $1$ ($\tfrac{1}{2}^-$) & $0.84^{+0.02}_{-0.02}$ & $(4.3^{+0.7}_{-0.6})^V$ & $2801.9^{+0.6}_{-0.7}$ & $\Sigma_c(2800)$ & $\sim 2800$ \\ 
$N D^*$ & $1$ ($\tfrac{1}{2}^-$) & $0.97^{+0.03}_{-0.02}$ & $(1.2^{+0.4}_{-0.4})^V$ & $2946.3^{+0.4}_{-0.4}$ & - & - \\
\hline\hline
$N \bar{D}^*$ & $0$ ($\tfrac{3}{2}^-$) & $0.94^{+0.01}_{-0.01}$ &
$(1.8^{+0.3}_{-0.2})^V$ & $2945.7^{+0.2}_{-0.3}$ & - & - \\
\hline \hline
$D \bar{D}$ & $0$ ($0^{++}$) & $0.63^{+0.08}_{-0.07}$  & $(1.5^{+3.5}_{-1.5})^V$ & $3733.0^{+1.5}_{-3.5}$ & - & - \\
$D^* \bar{D}$ & $0$ ($1^{++}$) & $0.89^{+0.20}_{-0.16}$  & $4.1^{+11.6}_{-4.1}$ & $3871.7^{+4.1}_{-11.6}$ & $X(3872)$ & $3871.69$ \\ 
$D^* \bar{D}^*$ & $0$ ($2^{++}$) & $0.93^{+0.20}_{-0.17}$  & $5.5^{+12.6}_{-5.8}$ & $4011.6^{+5.5}_{-12.6}$ & - & - \\
\hline \hline
$D_1 \bar{D}$ & $0$ ($1^{--}$) & $1.33^{+0.36}_{-0.31}$ & $34^{+30}_{-26}$ & $4255^{+26}_{-30}$ & $Y(4260)$ & $4218.6$ \\ 
$D_2 \bar{D}$ & $0$ ($2^{--}$) & $0.87^{+0.15}_{-0.13}$ & $2.7^{+7.3}_{-2.8}$ & $4325.6^{+2.8}_{-7.3}$ & - & - \\
$D_1 \bar{D}^*$ & $0$ ($1^{--}$) & $0.56^{+0.02}_{-0.02}$ & ${(4.1^{+1.1}_{-1.1})}^V$ & $4425.6^{+1.1}_{-1.1}$ & $Y(4360)$ & $4382.0$ \\
$D_2 \bar{D}^*$ & $0$ ($3^{--}$) & $1.89^{+0.60}_{-0.51}$ & $90^{+90}_{-73}$ & $4380^{+73}_{-90}$ & - & - \\
\hline  \hline
$D_1 \bar{D}_1$ & $0$ ($2^{++}$) & $1.66^{+0.47}_{-0.40}$ & $58^{+57}_{-45}$ & $4786^{+45}_{-57}$ & - & - \\
$D_2 \bar{D}_1$ & $0$ ($3^{+-}$) & $1.64^{+0.46}_{-0.39}$ & $56^{+55}_{-43}$ & $4829^{+43}_{-55}$ & - & - \\
$D_2 \bar{D}_1$ & $0$ ($3^{++}$) & $2.04^{+0.65}_{-0.58}$ & $97^{+95}_{-81}$ & $4788^{+81}_{-95}$ & - & - \\
$D_2 \bar{D}_2$ & $0$ ($3^{+-}$) & $0.83^{+0.11}_{-0.09}$ & $1.4^{+3.6}_{-1.5}$ &
$4924.7^{+1.4}_{-3.6}$ & - & - \\
$D_2 \bar{D}_2$ & $0$ ($4^{++}$) & $2.06^{+0.65}_{-0.54}$ & $98^{+96}_{-82}$ & $4828^{+82}_{-96}$ & - & - \\
\hline \hline
  $\Sigma_c \bar{D}$ & $\tfrac{1}{2}$ ($\tfrac{1}{2}^{-}$) & $1.00$ & $8.9$
& $4311.9$ & $P_c(4312)$ & $4311.9$ \\
  $\Sigma_c^* \bar{D}$ & $\tfrac{1}{2}$ ($\tfrac{3}{2}^{-}$) & $1.04$ & $9.4^{+1.7}_{-1.7}$
  & $4376.0^{+1.7}_{-1.7}$ & - & - \\
  $\Sigma_c \bar{D}^*$ & $\tfrac{1}{2}$ ($\tfrac{1}{2}^{-}$) & $0.85^{+0.06}_{-0.12}$ & $2.3^{+2.6}_{-1.9}$
& $4459.8^{+2.3}_{-2.9}$ & $P_c(4457)$ & $4457.3$ \\
$\Sigma_c \bar{D}^*$ & $\tfrac{1}{2}$ ($\tfrac{3}{2}^{-}$) & $1.13^{+0.04}_{-0.03}$ & $16.9^{+5.1}_{-4.7}$ & $4445.2^{+4.7}_{-5.1}$ & $P_c(4440)$  & $4440.3$ \\
  $\Sigma_c^* \bar{D}^*$ & $\tfrac{1}{2}$ ($\tfrac{1}{2}^{-}$) & $0.82^{+0.09}_{-0.04}$ & $1.3^{+2.7}_{-1.3}$
 & $4525.4^{+1.3}_{-2.7}$ & -  & - \\
$\Sigma_c^* \bar{D}^*$ & $\tfrac{1}{2}$ ($\tfrac{3}{2}^{-}$) & $0.96^{+0.03}_{-0.04}$ & $6.4^{+2.0}_{-2.0}$ & $4520.3^{+2.0}_{-2.0}$ & -  & - \\
$\Sigma_c^* \bar{D}^*$ & $\tfrac{1}{2}$ ($\tfrac{5}{2}^{-}$) & $1.19^{+0.06}_{-0.05}$ & $21.0^{+7.5}_{-6.8}$  & $4505.7^{+6.8}_{-7.5}$ & -  & - \\
\hline\hline
$\Sigma_c \bar{D}_1$ & $\tfrac{1}{2}$ ($\tfrac{1}{2}^{+}$) & $0.81^{+0.14}_{-0.15}$ & $1.0^{+3.8}_{-1.4}$ & $4874.6^{+1.0}_{-3.8}$ & -  & - \\
$\Sigma_c \bar{D}_1$ & $\tfrac{1}{2}$ ($\tfrac{3}{2}^{+}$) & $1.32^{+0.07}_{-0.06}$ & $29^{+12}_{-11}$ & $4847^{+11}_{-12}$ & -  & - \\
$\Sigma_c \bar{D}_2^*$ & $\tfrac{1}{2}$ ($\tfrac{3}{2}^{+}$) & $0.75^{+0.14}_{-0.18}$ & $0.1^{+3.3}_{-3.4}$
& $4916.5^{+0.1}_{-3.3}$ & -  & - \\
$\Sigma_c \bar{D}_2^*$ & $\tfrac{1}{2}$ ($\tfrac{5}{2}^{+}$) & $1.43^{+0.12}_{-0.10}$ & $37^{+18}_{-17}$ & $4879^{+17}_{-18}$ & -  & - \\
  \hline\hline
\end{tabular}
\caption{Selection of the molecular states predicted from saturation:
  ``System'' is the two-hadron system,
  $I(J^{P(C)})$ refers to the isospin, angular momentum,
  parity and C-parity (if applicable) of the state,
  $R_{\rm mol}$ is the relative interaction strength 
  with respect to the $P_c(4312)$ (see Eq.~(\ref{eq:R-mol})),
  $B_{\rm mol}$ the binding energy in MeV (where $(...)^V$ indicates
  a virtual state), $M_{\rm mol}$
  the mass of the molecule, ``Candidate'' refers to known resonances
  that might be identified with the predicted molecule and
  $M_{\rm candidate}$ is the candidate's mass (``$^1S_0$ pole''
  refers to the virtual state in singlet nucleon-nucleon scattering).
  The binding energies are calculated
  from Eqs.~(\ref{eq:bound}-\ref{eq:bound-R}),
  where for light-(heavy-) hadrons vertices we use
  the cutoff $\Lambda_L = 0.5\,{\rm GeV}$ ($\Lambda_H = 1.0\,{\rm GeV}$).
  The uncertainty in $R_{\rm mol}$ is obtained from varying the $m_{\sigma}$
  within the $400-550\,{\rm MeV}$ range, while for $B_{\rm mol}$
  we combine the previous uncertainty with a $\gamma_{\rm mol} / m_V$
  relative error by summing them in quadrature.
  For the hadron masses we use the isospin averages of
  the RPP values~\cite{Zyla:2020zbs}.
  The masses for the $\Lambda_c(2765)$, $\Lambda_c(2940)$, $\Sigma_c(2800)$
  and $X(3872)$ are taken from the RPP~\cite{Zyla:2020zbs} (we notice
  though that the $\Lambda_c(2765)$ is not 
  well-established 
  and could even be a $\Sigma_c$-type state or a superposition of
  two states instead, though Ref.~\cite{Belle:2019bab} considers
  it to be a $\Lambda_c$),
  for the $Y(4260)$ and $Y(4360)$
  we use the recent BESIII measurements~\cite{Ablikim:2020cyd} and
  for the $P_c(4312/4440/4457)$ we refer to the original
  LHCb observation~\cite{Aaij:2019vzc}.
}
\label{tab:predictions}
\end{table}

\parag{5. S-to-P-wave charmed meson transitions}
Now we want to explore the $Y(4260)$~\cite{Aubert:2005rm},
which has been conjectured to be a $J^{PC} = 1^{--}$ $D \bar{D}_1$
molecule~\cite{Liu:2005ay,Ding:2008gr,Wang:2013cya,Wang:2013kra,Cleven:2013mka,Chen:2019mgp},
though its nature remains unclear~\cite{Zhu:2005hp,LlanesEstrada:2005hz,Maiani:2005pe,Dubynskiy:2008mq,MartinezTorres:2009xb,Li:2013ssa}.

For the $D_{1(2)}^{(*)} \bar{D}^{(*)}$ two-hadron system the electric dipolar
\mpv{and magnetic quadrupolar} $D_{1(2)}^{(*)} \to D^{(*)}$ transitions
are possible, i.e. there are new $H_1 H_2 \to H_2 H_1$ components
in the potential not present in Eq.~(\ref{eq:V-cont}),
which we write as
\begin{eqnarray}
  V_C = C_0 + C_1\,\vec{\sigma}_{L1} \cdot \hat{\vec{S}}_{L2} +
  C_1' \, \vec{\Sigma}_{L1}^{\dagger} \cdot \vec{\Sigma}_{L2} +
  C_2' \, Q_{L1 ij}^{\dagger} Q_{L2 ij} \, , \nonumber \\
\end{eqnarray}
where $\vec{\sigma}_{L1}$ are the Pauli matrices
and $\hat{\vec{S}}_{L2}$ the spin-$\tfrac{3}{2}$
matrices as applied to the light-quark within the $D_{1(2)}^{(*)}$.
The dipolar and \mpv{quadrupolar pieces} are described by
the $C_1'$ and $C_2'$ couplings, 
$\vec{\Sigma}_L$ / $\vec{\Sigma}_L^\dagger$ are the spin matrices
for the $S_L = \tfrac{1}{2}$ to $\tfrac{3}{2}$ transition
(which can be consulted in \mpv{the appendices of}
Ref.~\cite{Lu:2017dvm}) \mpv{and
  $Q_{L ij} = (\sigma_{Li} \Sigma_{L j} + \sigma_{Lj} \Sigma_{Li})/2$.
  }
For saturating $C_1'$ and $C_2'$ we consider the Lagrangians
\begin{eqnarray}
  \mathcal{L}_{E1} &=& 
  \frac{f_{V}'}{2 M} \, {q_L}^{\dagger} \,
  {\Sigma}_{L i} \, \left( \partial_i V_0 - \partial_0 V_i \right)
  \, q_{L}' + {\rm C.C.}\, , \label{eq:vector-E1} \\
  \mathcal{L}_{M2} &=& 
  \frac{h_{V}'}{(2 M)^2} \, {q_L}^{\dagger} \,
  {Q}_{L i j} \,\partial_i \left( \epsilon_{j l m} \partial_l V_m \right)
  \, q_{L}' + {\rm C.C.}\, , \label{eq:vector-M2}
\end{eqnarray}
where $q_L$ and $q_L'$ are the non-relativistic light-quark subfields
for the S- and P-wave charmed mesons, generating the potentials
\begin{eqnarray}
  V_{E1} &=& -\frac{{f_V}'^2}{4 M^2}
    \frac{\omega_V^2 + \frac{1}{3} \mu_V^2}{{\vec{q}\,}^2 + \mu_V^2}\,
  \vec{\Sigma}_{L1}^{\dagger} \cdot \vec{\Sigma}_{L2} + \dots
  \, , \\
  V_{M2} &=& -\frac{{h_V}'^2}{16 M^4}\,\frac{1}{5}\,
  \frac{\mu_V^4}{{\vec{q}\,}^2 + \mu_V^2}\,
  Q_{L1 ij}^{\dagger} Q_{L2 ij} + \dots
\end{eqnarray}
where $\mu_V^2 = m_V^2 - \omega_V^2$ is the effective vector-meson mass
for this transition and $\omega_V = m(D_1) - m(D)$, $m(D_2^*) - m(D)$,
$m(D_1) - m(D^*)$ or $m(D_2^*) - m(D^*)$ for $D_1 \bar{D}$,
$D_2^* \bar{D}$, $D_1 \bar{D}^*$ or $D_2^* \bar{D}^*$
molecules.
The saturated couplings read
\begin{eqnarray}
  C_1^{'{\rm sat}}(m_V) &\propto& -\,\frac{{f_V'}^2}{4 M^2}\,
  {\left( \frac{m_V}{\mu_V} \right)}^{\alpha}\,\left( \frac{\omega_V^2 + \frac{1}{3}\mu_V^2}{\mu_V^2} \right)\,
  \left[ \zeta + \hat{\vec{I}}_1 \cdot \hat{\vec{I}}_2 \right] \, , \,\, \\
  C_2^{'{\rm sat}}(m_V) &\propto& -\,\frac{{h_V'}^2}{16 M^4}\,
  {\left( \frac{m_V}{\mu_V} \right)}^{\alpha}\,\frac{\mu_V^2}{5}\,
  \left[ \zeta + \hat{\vec{I}}_1 \cdot \hat{\vec{I}}_2 \right] \, , 
\end{eqnarray}
which includes the RGE correction derived in Eqs.~(\ref{eq:RGE}-\ref{eq:coupling-sat}).
If we define $f_V'=  \kappa_{E1}' \, g_V$, $\kappa_{E1}'$ can be
determined from  Eq.~(\ref{eq:vector-em-mixing}) and
the $D_{1(2)}^{(*)} \to D^{(*)} \gamma$ dipolar moment (extractable from
the partial decay widths~\cite{Close:2005se} or
the $\langle D_{1(2)}^{(*)} | r | D^{(*)} \rangle$
matrix elements ~\cite{Godfrey:2005ww}),
yielding $\kappa_{E1}' \sim (2.6-3.6)$.
This provides a fairly strong attraction in the $J^{PC} = 1^{--}$ $D \bar{D}_1$
\mpv{molecular configuration}, for which
$\vec{\Sigma}_{L1}^{\dagger} \cdot \vec{\Sigma}_{L2} = - 1$
(while $Q_{L1}^{\dagger} \cdot Q_{L2} \equiv Q_{L1 ij}^{\dagger} Q_{L2 ij} = 0$),
resulting in $B_{\rm mol} = 19-55\,{\rm MeV}$ and a mass of
$4235-4271\,{\rm MeV}$, to be compared with $4218.6 \pm 5.2$~\cite{Ablikim:2020cyd} for the $Y(4260)$ ($\kappa_{E1}'  \sim 3.9$ would reproduce the mass),
suggesting a sizable molecular component.

\mpv{
For the magnetic quadrupolar term, we define $h_V'=  \kappa_{M2}' \, g_V$,
where $\kappa_{M2}'$ can be extracted from $\kappa_{E1}'$ by the relation
$\kappa_{M2}' = (m_N/m_q)\,\kappa_{E1}' = 7.4-15.4$, with $m_N / m_q$
the ratio of the nucleon and constituent quark masses
in the particular quark-model calculation
used to obtain $\kappa_{M2}'$
(see Appendix~\ref{app:couplings} for details).
This provides a modest (but sizable) attraction
in the $J^{PC} = 1^{--}$ $D^* \bar{D}_1$
system (where $\vec{\Sigma}_{L1}^{\dagger} \cdot \vec{\Sigma}_{L2} = -\frac{1}{6}$
and $Q_{L1}^{\dagger} \cdot Q_{L2} = - \frac{5}{4}$), resulting in a virtual state
with $B_{\rm mol}^V = 4.1\,{\rm MeV}$ for $\kappa_{E1}' = 3.1$ and
$\kappa_{M2}' = 10.7$.
If we consider the uncertainties in the M1 and E2 couplings, the end result
ranges from a shallow bound state ($B_{\rm mol} = 1.1\,{\rm MeV}$)
to a virtual state moderately away from threshold
($B_{\rm mol}^V = 16.4\,{\rm MeV}$), while
for $\kappa_{E1}' = 3.9$ (which reproduces the $Y(4260)$) and
$\kappa_{M2}' = 24.6$, the location of the $Y(4360)$ as recently measured
by BESIII~\cite{Ablikim:2020cyd} would be reproduced.
Yet, the previous explanation does not consider the possible
$D^* \bar{D}_1$-$D^* \bar{D}_1^*$ coupled channel effects (the $D_1$ and
$D_1^*$ charmed mesons have about the same mass), which would make
binding more likely and explain the larger width of the $Y(4360)$,
check Appendix~\ref{subapp:CC-Y4360} for details.
Thus, while a pure molecular explanation of the $Y(4360)$ is less natural
than for the $Y(4260)$, a molecular component is nonetheless
possible and maybe even expectable. 
}

Other two interesting configurations are the $J^{PC} = 2^{--}$ $D \bar{D}_2^*$
and $3^{--}$ $D^* \bar{D}_2^*$ systems: the first depends on the coupling
$C_2'$ but not on $C_1'$ ($\vec{\Sigma}_{L1}^\dagger \cdot \vec{\Sigma}_{L2} = 0$
and $Q_{L1}^{\dagger} \cdot Q_{L2} = - 1$), which means that if observed
could be used to determine $\kappa_{M2}'$.
For $\kappa_{M2}' = 7.4-15.4$ we predict a binding energy and mass of
$B_{\rm mol} = (0.1-9.1)\,{\rm MeV}$ and $M = (4319.3-4327.5)\,{\rm MeV}$.
The second happens to be the most attractive configuration
($\vec{\Sigma}_{L1}^\dagger \cdot \vec{\Sigma}_{L2} = -1$
and $Q_{L1}^{\dagger} \cdot Q_{L2} = - \frac{1}{2}$),
with a state predicted somewhere in the
$M = (4349-4401)\,{\rm MeV}$ window.
In Table \ref{tab:predictions} we list the central value predictions
(i.e. $\kappa_{E1}' = 3.1$ and $\kappa_{M2}' = 10.7$) for
the four S- and P-wave molecules considered here.

\parag{6. Light baryons}

For baryons containing only light quarks,
the previous ideas can also be applied.
However there is a tweak which has to do with the relative sizes of
light-baryons in comparison with heavy-hadrons: light-hadrons
are larger than heavy-hadrons.
For instance, the electromagnetic radius $\sqrt{{\langle r^2 \rangle}_{\rm e.m.}}$
of the charged pion and kaon are about $0.66$ and $0.56\,{\rm fm}$,
respectively~\cite{Zyla:2020zbs}, with size decreasing once
the heavier strange quark is involved.
This pattern also applies to the charmed mesons, for which
$\sqrt{{\langle r^2 \rangle}_{\rm e.m.}} \sim (0.40-0.55)\,{\rm fm}$~\cite{Hwang:2001th,Becirevic:2009ya,Becirevic:2011cj,Can:2012tx}.
For baryons, the electromagnetic radius of the proton
is $0.84\,{\rm fm}$~\cite{Zyla:2020zbs}.
Lattice QCD calculations of the electromagnetic form factors of the singly
and doubly charmed baryons yield figures of
the order of $0.5$ and $0.4\,{\rm fm}$
respectively~\cite{Can:2013tna,Can:2013zpa},
half the proton radius,
where it is curious to notice that the doubly charmed baryons are
about the same size as the charmed mesons and only slightly smaller
than the singly charmed ones (from which the hypothesis
of using the same cutoff for all heavy hadrons
seems a sensible choice).

Of course, the problem is how to take this effect into account.
The easiest idea is to use a softer physical cutoff \mpv{(which effectively
  amounts to the introduction of a new parameter)} in the light baryon sector
\begin{eqnarray}
  \langle p' | V_C | p \rangle = C^{\rm sat}_{\rm mol}\,
  f(\frac{p'}{\Lambda_{\rm L}})\,f(\frac{p}{\Lambda_{\rm L}}) \, ,
\end{eqnarray}
where $\Lambda_{\rm L}$ will be close to $\Lambda_{H} / 2$, with
$C^{\rm sat}_{\rm mol}$ still determined as a ratio of
$C^{\rm sat}_{P_c}$.
For nucleons $g_S = 10.2$, $\tfrac{1}{3} g_{\omega} = g_{\rho} = g$
and $\kappa_{\rho} = 3.7$, $\kappa_{\omega} = -0.1$~\cite{Cordon:2009pj},
from which we get
\begin{eqnarray}
  R_{\rm singlet} = 1.77 \quad , \quad R_{\rm triplet} = 1.86 \, ,
\end{eqnarray}
which are really similar, reproducing Wigner's SU(4)
symmetry~\cite{Wigner:1936dx,Mehen:1999qs,Chen:2004rq}.
For $\Lambda_{L} = 0.5\,{\rm GeV}$ we predict shallow singlet/triplet
bound states
\begin{eqnarray}
  B_{\rm singlet} = 0.20 \,{\rm MeV} \quad , \quad
  B_{\rm triplet} = 0.94 \, {\rm MeV} \, ,
\end{eqnarray}
i.e. close to reality, where the singlet/triplet is a virtual/bound
state located at $0.07$/$2.22\,{\rm MeV}$ below threshold.
It is intriguing how reproducing the deuteron
forces us to choose a $\Lambda_L$ that basically coincides
with the scale at which Wigner's symmetry is expected to manifest~\cite{CalleCordon:2008cz,CalleCordon:2009ps,Timoteo:2011tt,RuizArriola:2016vap,Lee:2020esp}.
\mpv{
  Even though our choice of $\Lambda_L$ follows from a phenomenological
  argument, in practice this is a new parameter required for the correct
  description of systems containing a light baryon and it could have
  as well been determined from the condition of reproducing
  the deuteron or the virtual state.
}

Other system to consider is $\Delta \Delta$, with $g_{\rho} = g_{\omega} = 3 g$ and $\kappa_{\rho} = \kappa_{\omega} = \tfrac{3}{2}\,\mu_u$.
The most attractive configuration is $I=0$ and $S=3$ 
($B_{\Delta \Delta} = 15\,{\rm MeV}$ and $M_{\Delta \Delta} = 2405\,{\rm MeV}$ for
$M_{\Delta} \simeq 1210\,{\rm MeV}$~\cite{Zyla:2020zbs}),
which might be identified with the hexaquark predicted six decades
ago~\cite{Dyson:1964xwa}, the $d^*(2380)$ observed
in~\cite{Adlarson:2011bh} (which, however, has recently been argued to be
a triangle singularity~\cite{Molina:2021bwp})
or the $\Delta \Delta$ state computed
in the lattice~\cite{Gongyo:2020pyy}.

\parag{7. Light-heavy systems}

Finally for a two-hadron system with a light- and heavy-hadron, we simply
use different cutoffs for each hadron
\begin{eqnarray}
  \langle p' | V_C | p \rangle = C^{\rm sat}_{\rm mol}\,
  f(\frac{p'}{\Lambda_{\rm L}})\,f(\frac{p}{\Lambda_{\rm H}}) \, .
\end{eqnarray}
If we apply this idea to the $ND^{*}$ system, we find a bound state with
$I(J^P) = 0 (\tfrac{3}{2}^-)$ that might correspond to the $\Lambda_c(2940)$,
which has been theorized to be molecular~\cite{He:2006is,He:2010zq,Zhang:2012jk,Ortega:2012cx,Wang:2020dhf,Sakai:2020psu}.
\mpv{
  We also find it worth mentioning the prediction of a virtual state
  in the $ND$ system
with $I(J^P) = 1 (\tfrac{1}{2}^-)$, $B^V_{\rm mol} = 4.3\,{\rm MeV}$ and
$M = 2806.6\,{\rm MeV}$, which might be identified with the $\Sigma_c(2800)$
(also theorized to be molecular~\cite{Haidenbauer:2010ch,Sakai:2020psu}).
However, owing to the $\Sigma_c(2800)$ having being observed
in the $\Lambda_c \pi$ spectrum~\cite{Belle:2004zjl},
this interpretation is questionable unless it is only partially molecular
or there are other factors increasing the attraction
in the $I=1$ $ND$ system. We notice though that Ref.~\cite{Sakai:2020psu}
finds a $\Sigma_c(2800)$ resonance as a pole in $ND$ scattering.
In contrast, the $N\bar{D}^{(*)}$ system (i.e. singly charmed
pentaquark candidates) shows less attraction than the $ND^{(*)}$
owing to omega exchange becoming repulsive.
Yet, the $I(J^P) = 0 (\tfrac{3}{2}^-)$ $N \bar{D}^*$ configuration happens
to be close to binding, with a virtual state at
$B^V_{\rm mol} = 1.8\,{\rm MeV}$.
}

\parag{8. Conclusions}

We propose a description of heavy- (and light-)hadron molecules
in terms of an S-wave contact-range potential.
The couplings of this potential are determined from light-meson exchange
by means of a saturation procedure which incorporates a few RG ideas
to effectively combine together the contribution
from scalar and vector mesons.
In turn the light-meson exchange parameters are set from a series of well-known
phenomenological ideas and a cutoff $\Lambda_H \sim 1\,{\rm GeV}$ is included.
This procedure takes the $P_c(4312)$ as input, from which it is able to
reproduce the other two LHCb pentaquarks, the $X(3872)$ and predict
a few new molecular candidates.
If applied to the light-sector (with a few modifications),
it reproduces Wigner-SU(4) symmetry and
the deuteron as a shallow bound state.
Of course the question is whether the theoretical ideas contained
in this manuscript do really represent a good approximation to
the spectrum of molecular states.
Future experiments will tell, particularly the discovery of different spin
configurations of a given two-hadron system, as the hyperfine splittings
are very dependent on their origin, which we conjecture to be
the magnetic-like couplings of the vector mesons.

\vspace{0.15cm}

{\it Note added.} --- After the acceptance of this manuscript, the ALICE collaboration has presented the first experimental study of the $N\bar{D}$ interaction~\cite{ALICE:2022enj}. They extract the isoscalar ($I=0$) $N\bar{D}$ inverse scattering length ($f_0^{-1}$ in their notation), resulting in $f_0^{-1}(I=0) \in [-0.4,0.9]\,{\rm fm^{-1}}$. Our own calculation yields $f_0^{-1}(I=0) = 0.64^{+0.19}_{-0.15}\,{\rm fm}^{-1}$ (indicating the presence of a virtual state), or
$[0.49,0.83]\,{\rm fm}^{-1}$, a range which falls
within the ALICE estimation.

\section*{Acknowledgments}

We thank Feng-Kun Guo, Enrique Ruiz Arriola and Eulogio Oset
for their comments on this manuscript.
M.P.V. thanks the IJCLab of Orsay, where part of this work was done,
for its hospitality.
This work is partly supported by the National Natural Science Foundation
of China under Grants No. 11375024, the Fundamental
Research Funds for the Central Universities and the
Thousand Talents Plan for Young Professionals.

\appendix
\section{Determination of the M1, E1 and M2 couplings}
\label{app:couplings}

Here we briefly explain how we derive the vector meson couplings of
higher polarity, beginning with the magnetic M1 $\kappa_V$ couplings.
These are given by $\kappa_V = \frac{3}{2}\,\mu_u(j,l) / \mu_N$ with
$\mu_u(j,l)$ the magnetic moment of a light u-quark with total
and orbital angular momentum $j$ and $l$ (which also
characterize its parent heavy meson).
Within the quark model we expect the light-quark magnetic moment
operator to be
\begin{eqnarray}
  \hat{\mu}_q &=&
  \frac{e_q}{2 m_q}\,\left[ \vec{\sigma}_q + \vec{l}_q \right] 
  = \mu_q\,\left[ \vec{\sigma}_q + \vec{l}_q \right] \, ,
  \label{eq:M1-operator}
\end{eqnarray}
where $\mu_q$ is the magnetic moment of the $q = u,d$ quarks
in the quark model ($\mu_u \simeq 1.9\,\mu_N$ and $\mu_d \simeq -0.9\,\mu_N$),
$\vec{\sigma}_q$ the Pauli matrices as applied to the intrinsic spin
of the light-quark and $\vec{l}_q$ the light-quark orbital
angular momentum operator.
For a light-quark with $j$ and $l$ quantum numbers, its magnetic moment
is given by the matrix element
\begin{eqnarray}
  \mu_q(j,l) &=& \langle (s l) j j | \hat{\mu}_q | (s l) j j \rangle \, ,
\end{eqnarray}
with $| (s l) j j \rangle$ a state in which a light-quark with spin
$s = \tfrac{1}{2}$ and orbital angular momentum $l$ couples to
total angular momentum $j$ and third component $j$.
The calculation of this matrix element is trivial, yielding
\begin{eqnarray}
  \mu_q(j = l+\frac{1}{2},l) &=& \mu_q\,(j+\tfrac{1}{2}) \, , \\
  \mu_q(j = l-\frac{1}{2},l) &=& \mu_q\,\frac{j}{j+1}\,(j+\tfrac{1}{2}) \, , 
\end{eqnarray}
which translates into $\mu_u(\frac{1}{2},1) = \mu_u/3$ and
$\mu_u(\frac{3}{2},1) = 2 \mu_u$ for the $D_0$/$D_1^*$ and
$D_1$/$D_2^*$ P-wave charmed mesons.

The E1 and M2 couplings $\kappa_{E1}'$ and $\kappa_{M2}'$, which determine
the strength of the P-to-S-wave charmed meson transitions,
can be determined in turn from the electric dipolar and
magnetic quadrupolar moments of these transitions.
A comparison with the E1 and M2 electromagnetic Lagrangians
\begin{eqnarray}
  \mathcal{L}_{E1}
  &=& \langle e_q \rangle\,d_E'\,\, q_L^{\dagger}\,
  \vec{\Sigma}_L \cdot \left( \partial_0 \vec{A} - \vec{\partial} A_0 \right)
  \,q_L' + {\rm C.C.} \, , \\
  \mathcal{L}_{M2}
  &=& Q_M^q \, q_L^{\dagger}\, Q_{L ij} \partial_i B_j \, q_L' + {\rm C.C.} \, ,
\end{eqnarray}
together with Eq.~(\ref{eq:vector-em-mixing}) yields
$\kappa_{E1}' = 2 M \,\frac{3}{2}\,\langle e_u \rangle\,d_{E}'$
and $\kappa_{M2}' = (2 M)^2\,\frac{3}{2}\, Q_M^u$, where
in the Lagrangians above $q_L$, $q_L'$, $\vec{\Sigma}_L$
and $Q_{L ij}$ are defined as
below Eqs.~(\ref{eq:vector-E1}) and (\ref{eq:vector-M2}),
$\langle e_{q} \rangle = (m_Q e_q - m_q e_{\bar Q})/(m_q + m_Q)$ is
the effective charge of the $q \bar{Q}$ system
($\langle e_u \rangle = \frac{2}{3}$ for charmed antimesons),
$Q_M^q$ is the quadrupolar magnetic moment for the light-quark $q = u,d$,
$A_{\mu} = ( A_0, \vec{A} )$ the photon field and
$B_j = \epsilon_{j l m}\,\partial_l A_m$
the magnetic field.

If we begin with the E1 transitions, $\langle e_q \rangle\,d_E'$ is
the electric dipolar moment of the u-quark in the $D_{1(2)}^{(*)} \to D^{(*)}$
transition.
The dipolar moment can in turn be obtained in two different ways: (i)
from the matrix elements of the dipolar moment operator or (ii)
from the $D_{1(2)}^* \to D^{(*)} \gamma$ decays.
In the first case we consider the operator
\begin{eqnarray}
  {\hat{d}_{E}}^{q} = \langle e_{q} \rangle\,\vec{r} \, , 
\end{eqnarray}
where we define the dipolar moment in relation with the matrix element
\begin{eqnarray}
  \langle D_{1(2)}^{(*)} | \hat{d}_E^q | D^{(*)} \rangle = 
  \langle e_{q} \rangle\,d_E'\,\vec{\Sigma}_L \, .
\end{eqnarray}
We calculate $d_E'$ from the quark-model wave functions of
the $S_L = \tfrac{1}{2}$ or $\tfrac{3}{2}$ charmed meson,
which can be expanded as
\begin{eqnarray}
  | D(S_L ; J M) \rangle &=&
  \sum_{M_L M_H} \,
  \Psi_{S_L M_L}(\vec{r})\, | S_H M_H \rangle \langle S_L M_L S_H M_H |
  J M \rangle \, , \nonumber \\ \\
  \Psi_{S_L M_L}(\vec{r})
  &=& \sum_{\mu_l \mu_s} \, \frac{u_{l}(r)}{r} Y_{l \mu_l}(\hat{r})
  | s \, \mu_s \rangle
  \langle l \, \mu_l s \, \mu_s | S_L M_L \rangle \, , \nonumber \\
\end{eqnarray}
where $J, M$ refer to the total angular momentum of the charmed meson
and its third component, $S_L, M_L$ and $S_H\,=\frac{1}{2})$,
$M_H$ to the light- and heavy-quark spin, $\Psi_{S_L M_L}$ the wave function of
the light-quark, $l$, $\mu_l$ and $s\,(=\frac{1}{2})$, $\mu_s$
the orbital and intrinsic angular momentum of
the light-quark, $u_{l}$ the reduced wave function,
$Y_{l \mu_l}$ a spherical harmonic and $| j m \rangle$
and $\langle j_1 m_1 j_2 m_2 | j m \rangle$ refer to spin wave functions
and Clebsch-Gordan coefficients.
After a few manipulations we arrive at
\begin{eqnarray}
  d_E' = -\frac{\langle P | r | S \rangle}{\sqrt{3}} \, ,
\end{eqnarray}
where
\begin{eqnarray}
  \langle P | r | S \rangle = \int_0^{\infty} dr \, u_P(r)\, r \, u_S(r) \, ,
\end{eqnarray}
with $u_S$ and $u_P$ the $l = 0,1$ reduced wave functions.
We find that $\kappa_{E1}' = - 2 m_N d_E'$ (for $M = m_N$ and charmed mesons,
i.e. $\langle e_u \rangle = -\frac{2}{3}$) and
from $\langle P | r | S \rangle = 2.367\,{\rm GeV}$
in Ref.~\cite{Godfrey:2005ww}, we obtain
$\kappa_{E1}' =  2.6$.

In the second case, we use the electromagnetic decays of the $D_{1(2)}^{(*)}$
charmed mesons, which are described by the non-relativistic amplitude
\begin{eqnarray}
  \mathcal{A}(D_{1(2)}^{(*)} \to D^{(*)} \gamma) =
  \langle e_{q} \rangle\,d_E'\,\vec{\Sigma}_L \cdot \left(
  \partial_0 \vec{A} - \vec{\partial} A_0 \right) \, ,
\end{eqnarray}
from which the $D_2^{*0} \to D^{*0} \gamma$ decay reads
\begin{eqnarray}
  \Gamma(D_2^{*0} \to D^{*0} \gamma) = \frac{4 \alpha}{3}\,
  \frac{m(D^{*0})}{m(D_2^{*0})}\,q^3\,|d_E'|^2\,\, , \label{eq:Gamma-dE}
\end{eqnarray}
with $\alpha$ the fine structure constant and $q$ the momentum of
the photon.
If we use this decay width as calculated in Ref.~\cite{Close:2005se}
($\Gamma = 895\,{\rm keV}$, $q = 410\,{\rm MeV}$) we will arrive at
$\kappa_{E1}' = 3.6$.

For the magnetic quadrupolar moment ($Q_M^q$), it can be obtained
from the matrix elements of the M2 operator~\cite{Raab75}
\begin{eqnarray}
  \hat{Q}_M^q = \frac{e_q}{2 m_q} \, \left[
    \frac{1}{2}(\sigma_{q i} r_j + \sigma_{q j} r_i) +
    \frac{2}{3}(l_{q i} r_j + l_{q j} r_i) \right] \, ,
\end{eqnarray}
with $\mu_q$, $\vec{\sigma}_q$ and $\vec{l}_q$ as defined below
Eq.~(\ref{eq:M1-operator}).
The matrix element of this operator will be proportional to $Q_M^q$
\begin{eqnarray}
  \langle D_{1(2)}^{(*)} | \hat{Q}_M^q | D^{(*)} \rangle = Q_M^q\,Q_{L ij} \, ,
\end{eqnarray}
from which we obtain
\begin{eqnarray}
  Q_M^q = - \frac{e_q}{2 m_q} \, \frac{\langle P | r | S \rangle}{\sqrt{3}}
  = \frac{e_q}{2 m_q}\,d_E'\, .
\end{eqnarray}
If we use the quark model calculations of Ref.~\cite{Godfrey:2005ww},
where $m_u = 0.22\,{\rm GeV}$ and
$\langle P | r | S \rangle = 2.367\,{\rm GeV}$,
we arrive to $\kappa_{M2}' = 10.3$.
Instead, if we determine $d_E'$ from Ref.~\cite{Close:2005se}
(where $m_u = 0.33\,{\rm GeV}$) as below Eq.~(\ref{eq:Gamma-dE}),
we obtain $\kappa_{M2}' = 11.1$.
In this case, these two determinations yield similar results (with the average
being $\kappa_{M2}' = 10.7$), but this agreement is probably fortuitous and
the uncertainties should be of the same relative size as for $\kappa_{E1}'$.
Indeed, if we simply rewrite $\kappa_{M2}'= (m_N/m_u)\,\kappa_{E1}'$ and
vary $m_u$ independently of $\kappa_{E1}'$, we will obtain instead
the $\kappa_{M2}' = 7.4-15.4$ window, which is more in line with
what to expect from $\kappa_{E1}'$.

\section{Difficulties with the light-meson exchange model}
\label{app:OBE}

Here we consider a few theoretical difficulties with the light-meson
exchange picture.
The first is the nature of the scalar meson, which is not a pure $q\bar{q}$
state and contains tetraquark and molecular components as well
(e.g. the $\sigma$ appears as a wide resonance in $\pi \pi$ scattering,
check the recent review~\cite{Pelaez:2015qba}
and references therein).
The part of the $\sigma$ which is expected to manifest at the scale we are
saturating the couplings (i.e. $\Lambda \sim m_V$) is the $q \bar{q}$,
with the tetraquark components playing a more important role
at longer distances and manifesting themselves
as the two-pion exchange potential.
Dealing with this issue actually requires to also consider the large width
of the $\sigma$ (check the discussion below),
yet the previous observation suggests
treating the $\sigma$ that appears in the meson exchange
model as a standard meson, though with properties
that might differ from the physical $\sigma$
(a point of view which is for instance followed in the meson theory of
nuclear forces~\cite{Machleidt:1987hj,Machleidt:1989tm}).

The second is the width of the scalar meson, which raises the issue
of how this affects its exchange potential.
Several solutions exists in the literature, of which we underline
the following three: (i) to treat the exchange $\sigma$ as a
narrow effective degree of freedom, where its mass and coupling
within light-meson exchange models are not necessarily
the ones corresponding to a physical
$\sigma$~\cite{Machleidt:1987hj,Machleidt:1989tm},
(ii) the two-pole approximation of Binstock and Bryan~\cite{Binstock:1971duy},
in which the integral of the $\sigma$ propagator over its mass distribution is
approximated as the sum of two narrow particles, one lighter and
one heavier than the physical $\sigma$ and (iii) the treatment
by Flambaum and Shuryak~\cite{Flambaum:2007xj},
in which the previous integral is approximated
as the sum of several contributions,  of which
the two most important ones are the one corresponding
to the $\sigma$ pole (equivalent to the exchange of a narrow $\sigma$,
but with a weaker coupling) and another corresponding to
the exchange of its decay products (two pion exchange).
Here we choose the effective $\sigma$ solution, which is the simplest
and the one originally adopted in the meson theory of nuclear forces.
Yet, we notice that the RG equation as applied to saturation actually
relates these three solutions, as a decrease in the mass of the sigma
increases its effective strength, while the presence of medium range
two-pion exchange effects can be in turn substituted by a stronger
sigma exchange.
Indeed this is what actually happens in the meson exchange theory of the nuclear
force, where models in which there is no two-pion exchange require a stronger
$\sigma$ coupling~\cite{Machleidt:1987hj,Machleidt:1989tm}
than models which include it~\cite{Stoks:1994wp}.

A third problem is the exchange of heavier light-mesons with the same
quantum numbers as the scalar and vector mesons (e.g. the $\sigma$
can mix with the $f_0(1370)$, scalar glueballs
and other $0^{++}$ mesons).
Again, RG-improved saturation indicates that the heavier light mesons
can be included via the formula
\begin{eqnarray}
  C^{\rm sat}(m_V) = \sum_M f_{\rm sup}\left( \frac{m_V}{m_M} \right)
  \, C^M(m_M) \, ,
\end{eqnarray}
where $M$ and $m_M$ denote a given meson and its mass, and $f_{\rm sup}$ is
a suppression (or, if $m_M < m_V$, enhancement) factor.
For mesons with a mass similar to the saturation scale
(i.e. $\Lambda \sim m_V$), the suppression factor is expected to be
$f_{\rm sup}(x) = x^{\alpha}$ with $\alpha = 1$, as previously explained
below Eq.~(\ref{eq:coupling-combo}).
However if the mass is dissimilar, this suppression factor should deviate
more and more from the previous ansatz.
Besides, $f_{\rm sup}(x) = x^{\alpha}$ is only valid for light-mesons
with a mass not too different to the vector mesons and
eventually the finite size of the hadrons has to be taken
into account, which will probably lead to a considerably
larger suppression factor.
The previous discussion suggests nonetheless that heavier light-mesons
can actually be accounted for by a redefinition of the effective
couplings of the $\sigma$, $\rho$ and $\omega$ mesons, though
the modifications are expected to be small owing to
the aforementioned suppression of heavier meson
contributions.
Owing to the phenomenological nature of the relations we have used to obtain
the couplings and the aforementioned suppression, we consider that
this redefinition is not necessary.

\section{Pion exchange effects}
\label{app:OPE}

Here we revisit the assumption that pion exchanges are a perturbative effect
for the two-hadron systems we are considering.
For this we will explicitly include the one pion exchange (OPE) potential
in a few selected molecules and calculate the binding energy shift
$\Delta B_{\rm mol}^{\rm OPE}$ that it entails.
As we will see, $\Delta B_{\rm mol}^{\rm OPE}$ lies in general within
the binding uncertainties we have previously calculated.

For including the OPE potential we will do as follows: as we are limiting
ourselves to the S-wave approximation, we will only consider
the spin-spin component of OPE, which is given by
\begin{eqnarray}
  V_{\rm OPE}(\vec{q}) = \zeta\,\hat{T}_{12} \, \hat{C}_{L12}\,
  \frac{g_1 g_2}{6 f_{\pi}^2}\frac{\mu_{\pi}^2}{\mu_{\pi}^2 + {\vec{q}\,}^2} +
  \dots \, ,
\end{eqnarray}
with $\mu_{\pi}$ the effective pion mass, i.e. $\mu_{\pi}^2 = m_{\pi}^2 - \Delta^2$,
where $m_{\pi} \simeq 138\,{\rm MeV}$ is the pion mass in the isospin symmetric
limit and $\Delta$ is the mass difference between the hadrons emitting
(or absorbing) the virtual pion in each of the vertices,
in case they are different (e.g. the $D^* \bar{D} \to D \bar{D}^*$ case,
check for instance Ref.~\cite{Valderrama:2012jv} for a more detailed
discussion); the dots have the same meaning as in Eq.~(\ref{eq:M1})
and $\hat{T}_{12}$, $\hat{C}_{L12}$ were already defined below
Eq.~(\ref{eq:coupling-sat}).
We project this potential into S-waves, yielding
\begin{eqnarray}
  \langle p' | V_{\rm OPE} | p \rangle = \zeta\,\hat{T}_{12} \, \hat{C}_{L12}\,
  \frac{g_1 g_2}{24 f_{\pi}^2} \frac{\mu_{\pi}^2}{p\, p'}\,
  \log{\left[ \frac{\mu_{\pi}^2 + (p+p')^2}{\mu_{\pi}^2 + (p-p')^2} \right]} \, .
  \nonumber \\
\end{eqnarray}
To obtain the molecular potential, we add OPE to the contact-range potential
and regularize
\begin{eqnarray}
  \langle p' | V_{\rm mol} | p \rangle = \left( C^{\rm sat}_{\rm mol}(\Lambda_H) +
  \langle p' | V_{\rm OPE} | p \rangle \right)\,f(\frac{p'}{\Lambda_H})
  f(\frac{p}{\Lambda_H}) \, , \nonumber \\
\end{eqnarray}
where $f(x)$ is the regulator function (specifically,
the Gaussian regulator $f(x) = e^{-x^2}$).
This potential is plugged into the bound state equation
\begin{eqnarray}
  \phi(k) + 2\mu_{\rm mol}\,\int_0^{\infty} \frac{p^2\,dp}{2 \pi^2}
  \frac{\langle k | V_{\rm mol} | p \rangle}{p^2 + \gamma_{\rm mol}^2}
  \,\phi(p) = 0 \, ,
\end{eqnarray}
where, contrary to the purely contact-range case, the previous equation
cannot be solved analytically or semi-analytically when OPE is included.
The solution is obtained numerically instead by discretizing
the bound state equation, after which it becomes a linear system
that can be solved by standard means, where $\gamma_{\rm mol}$ is
calculated by finding the zeros of the determinant of
the matrix representing the linear system.

If we now define
\begin{eqnarray}
  \Delta B_{\rm mol}^{\rm OPE} = B_{\rm mol}^{\rm OPE} - B_{\rm mol} \, ,
\end{eqnarray}
for the $D^* \bar{D}$ and $D^* \bar{D}^*$ systems ($g_1 = g_2 = 0.6$) we obtain
\begin{eqnarray}
  \Delta B_{\rm mol}^{\rm OPE} &\approx& +0.0\,{\rm MeV} \quad
  \mbox{for $1^{++}$ $D^* \bar{D}$} \, , \\
  \Delta B_{\rm mol}^{\rm OPE} &=& -1.8\,{\rm MeV} \quad
  \mbox{for $2^{++}$ $D^* \bar{D}^*$} \, ,
\end{eqnarray}
which lies within the uncertainties we already have and
where for the $D^* \bar{D}$ system we have approximated
the effective pion mass to zero as
$m(D^*) - m(D) \approx m_{\pi}$.
For the $\Sigma_c \bar{D}^*$ and $\Sigma_c^* \bar{D}^*$ systems
($g_1 = 0.84$, $g_2 = 0.6$)
we obtain
\begin{eqnarray}
  \Delta B_{\rm mol}^{\rm OPE} &=& -1.6\,{\rm MeV} \quad
  \mbox{for $\tfrac{1}{2}^{-}$ $\bar{D}^* \Sigma_c$} \, , \\
  \Delta B_{\rm mol}^{\rm OPE} &=& +1.6\,{\rm MeV} \quad
  \mbox{for $\tfrac{3}{2}^{-}$ $\bar{D}^* \Sigma_c$} \, , \\
  \nonumber \\
    \Delta B_{\rm mol}^{\rm OPE} &=& -1.3\,{\rm MeV} \quad
  \mbox{for $\tfrac{1}{2}^{-}$ $\bar{D}^* \Sigma_c^*$} \, , \\
  \Delta B_{\rm mol}^{\rm OPE} &=& -1.2\,{\rm MeV} \quad
  \mbox{for $\tfrac{3}{2}^{-}$ $\bar{D}^* \Sigma_c^*$} \, , \\
  \Delta B_{\rm mol}^{\rm OPE} &=& +2.5\,{\rm MeV} \quad
  \mbox{for $\tfrac{5}{2}^{-}$ $\bar{D}^* \Sigma_c^*$} \, ,
\end{eqnarray}
which again lies within the estimated uncertainties of the model.
Finally, for the $NN$ system ($g_1 = g_2 = 1.29$) the effect of OPE
happens to be larger
\begin{eqnarray}
  \Delta B_{\rm mol}^{\rm OPE} &=& +1.8\,{\rm MeV} \quad
  \mbox{for the $^1S_0$ channel} \, , \\
  \Delta B_{\rm mol}^{\rm OPE} &=& +2.5\,{\rm MeV} \quad
  \mbox{for the deuteron} \, ,
\end{eqnarray}
which is about twice the size of the uncertainties we previously
estimated for these two systems.
This suggests that the present model could be improved in the light baryon
sector by explicitly including OPE in the future.

\section{Coupled channel effects}
\label{app:CC}

Here we consider a few examples of how coupled channel effects might affect
the predictions we have made.
The selected systems are (i) $\Sigma_c\bar{D}^*$-$\Sigma_c^* \bar{D}^*$,
(ii) $\Xi_c' \bar{D}$-$\Xi_c \bar{D}^* $ and
$\Xi_c \bar{D}^*$-$ \Xi_c^* \bar{D}$,
(iii) $D\bar{D}$-$D_s\bar{D}_s$ and (iv) $D^*\bar{D}_1$-$D^*\bar{D}_1^*$.
In general, these effects are smaller than the uncertainties we have already
estimated.
Yet, there might be exceptions in which coupled channel effects
could play an important role.

\subsection{The $\Sigma_c\bar{D}^*$-$\Sigma_c^* \bar{D}^*$ states}

We begin with the $P_c(4440)$ and $P_c(4457)$ pentaquark states,
which in our molecular model are $J=\tfrac{3}{2}$ and
$\tfrac{1}{2}$ $\Sigma_c \bar{D}^*$ bound states.
It happens that the $\Sigma_c \bar{D}^*$ and $\Sigma_c^* \bar{D}^*$ thresholds
are close, where the mass gap is $64.6\,{\rm MeV}$, and thus it might be
sensible to explicitly check whether this coupled channel effect
could play a significant role in the description of the $P_c(4440/4457)$.
The mechanism by which the two channels mix is the M1 interaction term
i.e. $C_1\,\hat{\vec{S}}_{L1} \cdot \hat{\vec{S}}_{L2}$
in Eq.~(\ref{eq:V-cont}).
The evaluation of the spin-spin operator for
$\Sigma_c \bar{D}^*$-$\Sigma_c^* \bar{D}^*$ yields
\begin{eqnarray}
  \hat{\vec{S}}_{L1} \cdot \hat{\vec{S}}_{L2} &=&
  \begin{pmatrix}
    - \frac{4}{3} & -\frac{\sqrt{2}}{3} \\
    -\frac{\sqrt{2}}{3} & -\frac{5}{2}
  \end{pmatrix} \quad \mbox{for $J=\tfrac{1}{2}$} \, \\
  \mbox{and} \quad &&
  \begin{pmatrix}
    + \frac{2}{3} & -\frac{\sqrt{5}}{3} \\
    -\frac{\sqrt{5}}{3} & -\frac{2}{2}
  \end{pmatrix} \quad \mbox{for $J=\tfrac{3}{2}$} \, ,
\end{eqnarray}
from which we can solve the coupled channel version of
the bound state equation, resulting in
\begin{eqnarray}
  M(\Sigma_c \bar{D}^*, \tfrac{1}{2}) &=& 4459.6\,(4458.9)\,{\rm MeV} \, , \\
  M(\Sigma_c^* \bar{D}^*, \tfrac{1}{2}) &=& 4525.5 - i\,0.1\,(4525.4)\,{\rm MeV} \, , \\
  \nonumber \\
  M(\Sigma_c \bar{D}^*, \tfrac{3}{2}) &=& 4444.3\,(4445.2)\,{\rm MeV} \, , \\
  M(\Sigma_c^* \bar{D}^*, \tfrac{3}{2}) &=& 4520.7 - i\,0.4\,(4520.2)
  \,{\rm MeV} \, .
\end{eqnarray}
These masses are close to the central value of the single channel
calculation (i.e. the values in parentheses), from which we are
driven to the conclusion that coupled channel effects are small
in the pentaquark case.

\subsection{The $\Xi_c' \bar{D}$-$\Xi_c \bar{D}^*$ and $\Xi_c \bar{D}^*$-$\Xi_c^* \bar{D}$ states}
\label{subapp:CC-Pcs4459}

A second example is the $P_{cs}(4459)$ pentaquark, which in the single
channel approximation is considered a $\Xi_c \bar{D}^*$ bound state.
Owing to the strange content of the $\Xi_c$ charmed baryon,
it is not clear what its coupling to the scalar meson is.
On the one hand, the naive expectation is that the $\sigma$ does not contain
a large $s\bar{s}$ component (if we assume it to be a $q\bar{q}$ state,
which is not clear to begin with).
If we combine this observation with the OZI rule, the coupling of
the $\sigma$ to the strange quarks within a baryon should be
smaller than to the $u$ and $d$ quarks.
On the other hand, the OZI rule is known to fail in the $0^{++}$
sector~~\cite{Geiger:1992va,Lipkin:1996ny,Isgur:2000ts,Meissner:2000bc},
which implies that the $\sigma$ should not be constrained by it.
From this we might expect a similar coupling for all the $u$, $d$ and $s$
light quarks.
We will adopt this second view, which implies $g_S = 6.8$ for the $\Xi_c$.
With this choice we obtain degenerate $J=\tfrac{1}{2}$, $\tfrac{3}{2}$
bound states with a mass of
\begin{eqnarray}
  M(\Xi_c \bar{D}^*) = 4466.9\,{\rm MeV} \, ,
\end{eqnarray}
which is to be compared with the experimental mass
$M(P_{cs}) = 4458.8 \pm 2.0\,{}^{+4.7}_{-1.1}\,{\rm MeV}$~\cite{Aaij:2020gdg}.

However, there are two nearby thresholds to be taken into account,
the $\Xi_c' \bar{D}$ and $\Xi_c^* \bar{D}$ for the $J=\tfrac{1}{2}$ and
$\tfrac{3}{2}$ cases, respectively.
The $\Xi_c \to \Xi_c^{('/*)}$ transition can be described by the Lagrangian
\begin{eqnarray}
  \mathcal{L}_{M1} &=& \frac{f_V}{2 M}\,d_{L0}^{\dagger} \epsilon_{ijk}\,
  \epsilon_{Li}\,
  \partial_j V_k \,d_{L1} + {\rm C.C.} \, ,
\end{eqnarray}
where $d_{L0}$ and $d_{L1}$ are fields representing
the $S=0$ and $1$ light diquarks
within the $\Xi_c$ and $\Xi_c'$/$\Xi_c^*$ baryons, respectively,
and $\vec{\epsilon}_L$ is the polarization vector of
the $S=1$ light diquark.
This Lagrangian can be matched to the electromagnetic one
\begin{eqnarray}
  \mathcal{L}_{M1} &=& \mu(1 \to 0)\,d_{L0}^{\dagger} \epsilon_{ijk}\,\epsilon_{Li}\,
  \partial_j A_k \,d_{L1} + {\rm C.C.} \, ,
\end{eqnarray}
with the diquark transition magnetic moment given by
$\mu(1 \to 0) = \mu(q_1) - \mu(q_2)$,
with $q_1 = u,d$ and $q_2 = s$
in the case at hand.
The actual $\Xi_c' \bar{D}$-$\Xi_c \bar{D}^*$ and
$\Xi_c \bar{D}^*$-$\Xi_c^* \bar{D}$ transitions
only involve the $\rho$ and $\omega$ vector mesons and thus we can ignore
the contribution from the strange meson to the magnetic moment
when applying vector meson dominance.
This leads to $\kappa_V = 2.9$ for the transition.
The saturation of the coupling for the transition yields
\begin{eqnarray}
  C^{\rm sat} = g_{V1} g_{V2}\,\left[ \zeta + \hat{T}_{12} \right]
  \,\frac{\kappa_{V1} \kappa_{V2}}{6 M^2}\,
  \vec{\epsilon}_{L1} \cdot \vec{\sigma}_{L2} \, ,
\end{eqnarray}
where the index $i =1,2$ represents the $\Xi_c$/$\Xi_c'$/$\Xi_c^*$ charmed
baryons and $\bar{D}^{(*)}$ charmed antimeson, respectively.
The evaluation of the light spin-spin operator yields
$| \vec{\epsilon}_{L1} \cdot \vec{\sigma}_{L2} | = 1$
for both $J=\tfrac{1}{2}$ and $\tfrac{3}{2}$.
From this, we solve the coupled channel bound state equation and obtain
\begin{eqnarray}
  M(\Xi_c \bar{D}^*, J=\tfrac{1}{2}) &=& 4468.4 - i\,1.4\,(4466.9)\,
  {\rm MeV} \, , \\
  M(\Xi_c \bar{D}^*, J=\tfrac{3}{2}) &=& 4464.4\,(4466.9)\,{\rm MeV} \, ,
\end{eqnarray}
where the value in parentheses is the previous single channel calculation.
In this latter case, the $J=\tfrac{3}{2}$ state is close to the experimental
single peak solution, but it also compares well with the two peak solution
considered in Ref.~\cite{Aaij:2020gdg},
$M_1 = 4454.9 \pm 2.7\,{\rm MeV}$ and
$M_2 = 4467.8 \pm 3.7\,{\rm MeV}$,
suggesting that the spin of the lower (higher) mass state
should be $J = \tfrac{3}{2}$ ($\tfrac{1}{2}$).

\subsection{The $D\bar{D}$-$D_s\bar{D}_s$ states}

A second example is the $D\bar{D}$ system, for which a bound state
has been predicted in the lattice~\cite{Prelovsek:2020eiw} with
$B_{\rm mol} = 4.0^{+5.0}_{-3.7}$.
Here we predict a virtual state instead (which could bind
within uncertainties), but we did not include the $D\bar{D}$-$D_s\bar{D}_s$
coupled channel dynamics of Ref.~\cite{Prelovsek:2020eiw}.
Thus it is worth the effort to explore the importance of this channel.

For the coupled channel dynamics, the $D\bar{D}$-$D_s\bar{D}_s$ transition
potential is given by vector meson ($K^*(890)$) exchange
\begin{eqnarray}
  && V(H\bar{H}-H_s\bar{H}_s) = \nonumber \\
  && \quad -2\sqrt{2}\,\frac{g_V^2}{q^2 + m_{K^*}^2}\,\left[
    1 + \kappa_V^2\frac{m_{K^*}^2}{6 M^2}\,
    \hat{\vec{S}}_{L1} \cdot \hat{\vec{S}}_{L2}
    \right] \, ,
\end{eqnarray}
with $H=D$,$D^*$ and $H_s=D_s$,$D_s^*$ and $V = K^*$, where $g_V$ and
$\kappa_V$ are the standard couplings for this system and
the $2\sqrt{2}$ factor originates from SU(3)-flavor
symmetry.
For the $D_s \bar{D}_s$ diagonal potential, it will be given by
scalar and vector meson ($\phi(1020)$) exchange
\begin{eqnarray}
  && V(H_s\bar{H}_s) = \nonumber \\
  && \quad -\frac{{g_S'}^2}{q^2 + m_S^2} -
  2\,\frac{g_V^2}{q^2 + m_{\phi}^2}\,\left[
    1 + \kappa_V^2\frac{m_{\phi}^2}{6 M^2}\,
    \hat{\vec{S}}_{L1} \cdot \hat{\vec{S}}_{L2}
    \right] \, , \nonumber \\
\end{eqnarray}
where $g_S'$ refers to the coupling of the $\sigma$ to the $D_s^{(*)}$ meson
and $m_{\phi} = 1020\,{\rm MeV}$ is the mass of the $\phi(1020)$.
Regarding $g_S'$ and the $D_s^{(*)}$, we follow the same line of argumentation
as for the $\Xi_c$ in Appendix~\ref{subapp:CC-Pcs4459} and take $g_S' = 3.4$.
The saturated couplings read
\begin{eqnarray}
  && C(H\bar{H}-H_s\bar{H}_s) =  \nonumber \\
  && \quad -2\sqrt{2}\,{\left( \frac{m_V}{m_{K^*}}\right)}^{\alpha}
  \frac{g_V^2}{m_{K^*}^2}\,\left[
    1 + \kappa_V^2\frac{m_{K^*}^2}{6 M^2}\,
    \hat{\vec{S}}_{L1} \cdot \hat{\vec{S}}_{L2}
    \right] \, , \\
  && C(H_s\bar{H}_s) = \nonumber \\
  && \quad -\,{\left( \frac{m_V}{m_{S}}\right)}^{\alpha}\,\frac{{g_S'}^2}{m_S^2} -
  2\,\,{\left( \frac{m_V}{m_{\phi}}\right)}^{\alpha}\,
  \frac{g_V^2}{m_{\phi}^2}\,\left[
    1 + \kappa_V^2\frac{m_{\phi}^2}{6 M^2}\,
    \hat{\vec{S}}_{L1} \cdot \hat{\vec{S}}_{L2}
    \right] \, , \nonumber \\
\end{eqnarray}
and concrete calculations yield
\begin{eqnarray}
  M(D\bar{D}) &=& 3733.5^V \, (3733.0^V) \, {\rm MeV} \, , \\
  M(D_s\bar{D}_s) &=& 3928.8^V - i\,2.4^V \, (3929.8^V) \, {\rm MeV} \, , 
\end{eqnarray}
where the superscript $V$ indicates a virtual state solution and
the masses in parentheses represent the prediction of
the single channel calculation.
That is, this coupled channel effect indeed provides an attractive
contribution to the $D\bar{D}$ system (about half a ${\rm MeV}$),
but it is still small in comparison with the uncertainties
we have for the location of the $D\bar{D}$.

\subsection{The $D^*\bar{D}_1$-$D^*\bar{D}_1^*$ states}
\label{subapp:CC-Y4360}

The mass difference between the $D_1$ ($S_L = \tfrac{3}{2}$) and
$D_1^*$ ($S_L = \tfrac{1}{2}$) charmed mesons is small
\begin{eqnarray}
  m(D_1) - m(D_1^*) = (10 \pm 9) \,{\rm MeV} \, ,
\end{eqnarray}
where most of the uncertainty comes from the broad $D_1^*$ charmed meson.
As a consequence, if we try to explain the $Y(4360)$ as a $D^* \bar{D}_1$
molecule, it will be difficult not to consider too the possible mixing
with the $D^* \bar{D}_1^*$ system.
Indeed, if both the $Y(4260)$ and $Y(4360)$ were to be $D \bar{D}_1$ and
$D^* \bar{D}_1$ molecules, respectively, the possible mixing of
the $Y(4360)$ with the $D^* \bar{D}_1^*$ channel might very
well explain why it is considerably broader
than the $Y(4260)$.
In the following lines we will explain how to include
the $D^*\bar{D}_1$-$D^*\bar{D}_1^*$
coupled channel dynamics.

We begin with the M1 $D_1 \to D_1^*$ vector-meson transitions, for which
the Lagrangian reads
\begin{eqnarray}
  \mathcal{L}_{M1} &=& \frac{f_V}{2 M}\,{q_L''}^{\dagger} \epsilon_{ijk} \Sigma_{Li}
  \partial_j V_k \,q_L' + {\rm C.C.} \, ,
\end{eqnarray}
where $q_L'$ and $q_L''$ refer to the light-quark subfield
for the $S_L = \tfrac{3}{2}$ and $S_L = \tfrac{1}{2}$
P-wave charmed mesons respectively.
By matching with the electromagnetic Lagrangian
\begin{eqnarray}
  \mathcal{L}_{M1} &=& \mu_q(\tfrac{3}{2} \to \tfrac{1}{2})\,
          {q_L''}^{\dagger} \epsilon_{ijk} \Sigma_{Li}
          \partial_j A_k \,q_L' + {\rm C.C.} \, ,
\end{eqnarray}
we obtain $\kappa_V = \frac{3}{2} \mu_u(\tfrac{3}{2} \to \tfrac{1}{2})$, where
the transition magnetic moment can be extracted from the matrix elements of
the magnetic moment operator in Eq.~(\ref{eq:M1-operator}), yielding
$\mu_u(\tfrac{3}{2} \to \tfrac{1}{2}) = \mu_u / \sqrt{3}$.
The next is the E1 $D \to D_1^*$ vector-meson transition, which can be
obtained from the matrix elements of
\begin{eqnarray}
  \langle D_{0(1)}^{(*)} | \hat{d}_E^q | D^{(*)} \rangle = 
  \langle e_{q} \rangle\,d_E''\,\vec{\sigma}_L \, ,
\end{eqnarray}
with $d_E'' = -\langle P | r | S \rangle / 3$, i.e. $1/\sqrt{3}$ smaller than
for the $D \to D_1$ family of transitions.
This give us $\kappa_{E1}'' = \kappa_{E1}' / \sqrt{3}$.

The saturated coupling for the $D^*\bar{D}_1$-$D^*\bar{D}_1^*$ transition reads
\begin{eqnarray}
  C^{\rm sat} \propto
  C_{M1}\,{\left( \vec{\sigma}_{L1} \cdot \vec{\Sigma}_{L2} \right)}_D +
  C_{E1}\,{\left( \vec{\sigma}_{L1} \cdot \vec{\Sigma}_{L2} \right)}_E \, , 
\end{eqnarray}
where the $D$ and $E$ subscripts stand for ``direct''
($D^*\bar{D}_1 \to D^*\bar{D}_1^*$) and ``exchange''
($D^*\bar{D}_1 \to {D}_1^*\bar{D}^*$) terms,
the difference being that the sign of the exchange term depends
on the C-parity of the system under consideration
(for $J^{PC} = 1^{--}$ the exchange and direct terms
have the same sign).
The $C_{M1}$ and $C_{E1}$ contributions read
\begin{eqnarray}
  C_{M1} &=& +g_{V}^2\,\left[ \zeta + \hat{T}_{12} \right]
  \,\frac{\kappa_{M1} \kappa_{M1}'}{6 M^2}\,
  \, , \\
  C_{E1} &=& -{g_{V}^2}\,\left[ \zeta + \hat{T}_{12} \right]\,
  \frac{\kappa_{E1}'\,\kappa_{E1}''}{4 M^2}
  {\left( \frac{m_V}{\mu_V} \right)}^{\alpha}\,
  \left( \frac{\omega^2 + \frac{1}{3}\mu_V^2}{\mu_V^2} \right)\,
  , \nonumber \\
\end{eqnarray}
where for $J^{PC} = 1^{--}$ the M1 and E1 terms end up interfering destructively,
leading to relatively weak coupled channel dynamics (despite
the two thresholds being so close).
If we use $M=m_N$, $\kappa_{M1} = \sqrt{3} \kappa_{M1}' = 2.9$ and
$\kappa_{E1}' = \sqrt{3} \kappa_{E1}'' \simeq 3.1$,
we end up with a weakly bound $D^* D_1$-$D^* D_1^*$ state with a mass
of $4420 \pm 9\,{\rm MeV}$.
However if we employ $\kappa_{E1}' = \sqrt{3} \kappa_{E1}'' \simeq 3.9$
(the value of the E1 coupling that reproduces the $Y(4260)$) and the higher
end value of the M2 coupling for this choice of $\kappa_{E1}'$,
i.e. $\kappa_{M2}' \simeq 16.7$, we will predict a bound state
with mass of $4417^{+6}_{-8}\,{\rm MeV}$.


%

\end{document}